\def\d{\mbox{d}}
\documentstyle[pra,aps,amssymb]{revtex}
\begin{document}
\draft

\wideabs{

\title{Dilute Bose gas in two dimensions:\\
Density expansions and the Gross-Pitaevskii equation}

\author{A. Yu. Cherny and A. A. Shanenko}

\address{Bogoliubov Laboratory of Theoretical Physics,
Joint Institute for Nuclear Research, 141980, Dubna, Moscow region,
Russia}

\date{February 28, 2001}

\maketitle

\begin{abstract}
A dilute two-dimensional (2D) Bose gas at zero temperature is studied
by the method developed earlier by the authors. Low density
expansions are derived for the chemical potential, ground state
energy, kinetic and interaction energies. The expansion parameter is
found to be a dimensionless in-medium scattering amplitude $u$
obeying the equation $1/u+\ln u=-\ln(na^{2}\pi)-2\gamma$, where
$na^{2}$ and $\gamma$ are the gas parameter and the Euler constant,
respectively. It is shown that the ground state energy is mostly
kinetic in the low density limit; this result does not depend on a
specific form of the pairwise interaction potential, contrary to
3D case. A new form of 2D Gross-Pitaevskii equation is proposed
within our scheme.
\end{abstract}


}

\narrowtext



Theoretical investigation of 2D Bose gas is of interest not only in
itself but also from the point of view of its experimental
applications. Indeed, the experimental observation of the
Bose-Einstein quasicondensate in the Bose gas of hydrogen atoms was
claimed~\cite{safonov}. On the other hand, the discovery of the
Bose-Einstein condensation in magnetically trapped alkali-metal
atoms~\cite{anderson} stimulated rapid progress in optical cooling
and trapping of atoms. This progress gives us hope that an
experimental observation of 2D quasicondensation in the trapped
atoms is a matter of near future (see some experiments along this
line in Ref.~\cite{glauck}). Theoretically, the Bose-Einstein
condensation is associated with the off-diagonal long-range order,
i.e., the non-zero asymptotic at $r=|{\bf r}_{1}-{\bf r}_{2}|\to\infty$
(from the physical point of view at $r\gg1/\sqrt{n}$)
for the one-body density matrix
\begin{equation}
\langle\hat{\psi}^{\dagger}({\bf r}_{1})\hat{\psi}({\bf r}_{2})
\rangle \to
\langle\hat{\psi}^{\dagger}({\bf r}_{1})\rangle\langle
\hat{\psi}({\bf r}_{2})\rangle\neq 0.
\label{ODLRO}
\end{equation}
Here $\hat{\psi}^{\dagger}({\bf r})$ and $\hat{\psi}({\bf r})$ are
the Bose field operators, $\langle\cdots\rangle$ stands for the
statistical average, and $\langle\hat{\psi}({\bf r})\rangle=\phi({\bf
r})$ is the order parameter. As was shown by Hohenberg~\cite{hoh}
from the Bogoliubov ``$1/q^{2}$" theorem~\cite{bog}, in 2D case there
is no off-diagonal long-range order at finite temperatures due to the
temperature long-range fluctuations of the phase, and the limit
(\ref{ODLRO}) is equal to zero.  In spite of this fact, a phase
transition is possible to a superfluid state at sufficiently low
temperature $T_{c}$~\cite{berez}. At this temperature, the asymptotic
behaviour of the one-body density matrix at $r\to\infty$ is changed
from an exponential decay (above
$T_{c}$) to a power decay (below $T_{c}$) with respect to $r$.  Thus
one can speak about the phenomenon of the quasicondensation in two
dimensions. For zero temperature, the limit (\ref{ODLRO}) differs
from zero, and, hence, there exists the true Bose-Einstein
condensate. Assuming that the condensate does exist at $T=0$, in this
paper we consider low-density expansions for 2D homogeneous Bose
gas with respect to the gas parameter $na^{2}$, where $n=N/S$ is the
density (a number of particles per unit area), and $a$ stands for the
{\it two-dimensional} scattering length (see Appendix~\ref{scat}).
Note that the density expansion for the chemical potential is
intimately related to a form of the Gross-Pitaevskii equation, a
powerful tool for investigating a dilute inhomogeneous system of Bose
particles~\cite{dalf}. As the chemical potential is a continuous
function of temperature, one can expect that our results, obtained
for zero temperature, are valid also for finite temperatures $T\ll
T_{c}$. Below we consider the case of zero temperature only.

The leading term of the energy expansion in $na^2$ for a gas of hard
discs was first obtained by Schick~\cite{schick}, who made use of the
Beliaev method~\cite{bel}, developed for 3D Bose gas. Recently the
Schick asymptotic formula has been proved rigorously~\cite{lieb}. To
the best of our knowledge, there is only one paper, by Hines {\it et
al.}~\cite{hines}, where the next-to-leading terms were evaluated.
However, the authors of Ref.~\cite{hines} employed the {\it
first-order} Beliaev approximation for obtaining the {\it
next-to-leading} terms, while in three dimensions the {\it
second-order} Beliaev approximation is needed for the same purpose.
Following Schick, they ignored the imaginary part of the Beliaev
equation for the chemical potential (compare Eq.~(4.3) in
Ref.~\cite{bel} with Eq.~(1) in Ref.~\cite{hines}). It is the
unphysical imaginary correction to the chemical potential that
determines the range of validity of the first-order approximation
itself. One can easily demonstrate that in 2D case the correction is
of the order of $n/\ln^{2}(na^{2})$; therefore, that method can not
yield correct terms of the expansion in this order and higher. At the
same time, our method~\cite{our} successfully reproduces in 3D case
the famous next-to-leading term for the chemical potential, which
ensures that our results are valid also in two dimensions. Thus it
becomes clear why only the first two terms involved in our expansion
[see Eqs.~(\ref{udelta}), (\ref{eexpdelta}) below] coincide with
those of the corresponding expansion from Ref.~\cite{hines}: the
correct coefficient even for the third term is beyond the
approximation used in Ref.~\cite{hines}.

In this paper we adopt the method developed and described in detail
in our previous publications~\cite{our}. Only basic notations and
some important points are discussed here. For a homogeneous system,
the one-body density matrix $\langle\hat{\psi}^{\dagger}({\bf
r}_{1})\hat{\psi}({\bf r}_{2})\rangle$ depends on ${\bf r}={\bf
r}_{1} - {\bf r}_{2}$, and, hence, its eigenfunctions and eigenvalues
are the plane waves $\exp(i{\bf p}\cdot{\bf r})/\sqrt{S}$ and the
occupation numbers $n_{p}=\langle \hat{a}_{\bf p}^{\dagger}
\hat{a}_{\bf p} \rangle$, respectively. The Bose-Einstein condensate
corresponds to the macroscopic occupation number $N_{0}$, and the
order parameter is $\langle\hat{\psi}({\bf r})\rangle =\langle
\hat{a}_{0} \rangle/\sqrt{S} = \sqrt{n_{0}} e^{i\chi}$ ($n_{0}$
stands for the density of the condensate). In turn, the
eigenfunctions of the two-body density matrix $\langle
\hat{\psi}^{\dagger} ({\bf r}_{1}) \hat{\psi}^{\dagger}({\bf r}_{2})
\hat{\psi}({\bf r}'_{2}) \hat{\psi}({\bf r}'_{1})\rangle$ can be
naturally classified as follows. The maximum eigenvalue
$N_{0}(N_{0}-1)\sim N_{0}^{2}$ corresponds to the state of two
particles in the condensate; its eigenfunction $\varphi(r)/S$ can be
interpreted as a pair wave function {\it in medium} of the
condensate-condensate type. The other macroscopic eigenvalues
$2N_{0}n_{q}$ correspond to the two-body states with one particle in
the condensate and another one beyond the condensate; its
eigenfunctions $\varphi_{{\bf q}/2}({\bf r})\exp[i{\bf q}\cdot({\bf
r}_{1}+{\bf r}_{2})/2]/S$ are of the condensate-supracondensate type,
where $\hbar {\bf q}$ is the total momentum of the pair of bosons.
The residuary non-macroscopic eigenvalues are related to the
supracondensate-supracondensate pairs and to bound ones provided the
latter exist.  The functions $\varphi(r)$ and $\varphi_{{\bf
q}/2}({\bf r})$ can be chosen as real quantities given by ($p\not=0$)
\begin{equation}
\varphi(r)=1+\psi(r),\;
\varphi_{{\bf p}}({\bf r})= \sqrt{2}\cos({\bf p}\cdot{\bf r})+
\psi_{{\bf p}}({\bf r})
\label{phi}
\end{equation}
with the boundary conditions $\psi(r),\;\psi_{{\bf p}} ({\bf r}) \to
0$ at $r \to \infty$.  The Fourier transforms of the scattering parts
can be expressed in terms of the Bose operators:
\begin{eqnarray}
\psi(k)=
\frac{\langle \hat{a}_{{\bf k}}\hat{a}_{-{\bf k}}\rangle}{n_0},\
{\psi}_{\bf p}({\bf k})= \sqrt{\frac{S}{2 n_0}}
\frac{\langle \hat{a}^{\dagger}_{2{\bf p}}
\hat{a}_{{\bf p}+{\bf k}} \hat{a}_{{\bf p}-{\bf k}}\rangle}{n_{2p}}.
\label{psik}
\end{eqnarray}
With the help of the in-medium scattering amplitudes
$U(k) = \int\d^2r\,\varphi(r)V(r) \exp[-i{\bf k}\cdot{\bf r}]$ and
$U_{\bf p}({\bf k}) = \int\d^2r\,\varphi_{\bf p}({\bf r})V(r)
 \exp[-i{\bf k}\cdot{\bf r}]$,
the chemical potential reads
\begin{equation}
\mu=n_0 {U}(0)
+\sqrt{2}\int\frac{\d^2q}{(2\pi)^2}n_q{U}_{{\bf q}/2}({\bf q}/2).
\label{mu}
\end{equation}
Here we introduce a pairwise interaction potential $V(r)$. It should
be emphasized that the formulas (\ref{psik}) and (\ref{mu}),
derived~\cite{our} within the Bogoliubov principle of the correlation
weakening~\cite{bog}, are {\it exact}.  For a Bose gas, a system with
a small condensate depletion $(n-n_0)/n \ll 1$, the pair distribution
function is expressed as
\begin{equation}
g(r)=\left(\frac{n_0}{n}\right)^2\varphi^2(r)+2\frac{n_0}{n}
\int\frac{\d^2q}{(2\pi)^2}\frac{n_q}{n}\varphi^2_{{\bf q }/2}({\bf r}),
\label{gr}
\end{equation}
where the contribution of the supracondensate-sup\-ra\-con\-den\-sate
pair wave functions can be neglected.  Another restriction for this
representation is the assumption that there are no bound pair
states~\cite{our}. In order to fulfil the latter condition, it is
sufficient to require $V(r)>0$, and, as usually, $V(r)\to 0$ for
$r\to\infty$. In the framework of our scheme, the following equations
are valid {\it at sufficiently low densities}~\cite{our}:
\begin{eqnarray}
n_{k}&=&\frac{1}{2}\Biggl(\frac{T_k+nU(k)}{\sqrt{T_k^2+2nT_k U(k)}}
-1\Biggr),\label{nk} \\
\psi(k)&=&-\frac{1}{2}\frac{U(k)}{\sqrt{T_k^2+2nT_kU(k)}},
\label{psikU}
\end{eqnarray}
with $T_{k}=\hbar^{2}k^{2}/(2m)$. Equations (\ref{nk}) and
(\ref{psikU}) look like those of the modified Bogoliubov model where
the ``bare" pairwise potential $V(k)$ is replaced by the effective
one $U(k)$ that is determined from the two-body Schr\"odinger equation.
However, in our method there exists a key difference, which is of
particular importance in two dimensions:  Eq.~(\ref{psikU}) is a {\it
self-consistent} equation for the in-medium scattering amplitude
$U(k)$. Indeed, using the definition of the latter, Eq.~(\ref{psikU})
can be represented in the Lippmann-Schwinger form
\begin{equation}
U(k)=V(k)-\frac{1}{2}\int\frac{\d^2q}{(2\pi)^2}
\frac{V(|{\bf k}-{\bf q}|)U(q)}{\sqrt{T^2_q+2nT_qU(q)}}.
\label{Uk}
\end{equation}
Besides, one can make use of the limiting relation [see Eq. (\ref{phi})]
$\lim_{p \to 0} \varphi_{{\bf p}} ({\bf r})=\sqrt{2}\varphi(r)$, which
allows one to simplify Eqs.~(\ref{mu}) and (\ref{gr})
\begin{eqnarray}
\mu&=&nU(0)[1+(n-n_0)/n+\cdots],
\label{muasymp} \\
g(r)&=&\varphi^{2}(r)[1+2(n-n_0)/n+\cdots].
\label{grasymp}
\end{eqnarray}
Thus, our scheme is reduced to the following. First, one should solve
Eq.~(\ref{Uk}) and find $U(0)$ [and $\varphi(r)$] as a function of
the density at $n\to0$.  Second, the condensate depletion $(n -
n_{0}) / n$ should be determined from Eq.~(\ref{nk}). Third,
employing these results, one should obtain the density expansion for
the chemical potential (\ref{muasymp}) and the short-range behaviour
for the pair distribution function (\ref{grasymp}). Note that in
three dimensions this scheme is in excellent agreement with the data
of Monte-Carlo calculations for hard spheres (see the last paper in
Ref.~\cite{our}).

In order to solve Eq.~(\ref{Uk}) at $n \to 0$, we employ the
procedure of linearization, which is similar to that in 3D
case~\cite{our}, and rewrite this equation in the form
\begin{equation}
U(k)=V(k)-\frac{1}{2}\mbox{P.P.}\!\int\!\frac{\d^2q}{(2\pi)^2}
     \frac{V(|{\bf k}-{\bf q}|)U(q)}{T_{q}-T_{q_{0}}}-\frac{I}{2},
\label{Uk1}
\end{equation}
where P.P. denotes the Cauchy principal value, and  $I$ is given by
\[
I=\mbox{P.P.}\int\frac{\d^2q}{(2\pi)^2}\Biggl[
  \frac{V(|{\bf k}-{\bf q}|)U(q)}{\sqrt{T^2_q+ 2nT_qU(q)}}-
  \frac{V(|{\bf k}-{\bf q}|){U}(q)}{T_{q}-T_{q_{0}}}\Biggr].
\]
Here we introduce the auxiliary quantity $q_{0} =c \sqrt{2mnU(0)}
/\hbar$, where $c$ stands for an arbitrary dimensionless constant.
Performing the ``scaling" substitution
\begin{equation}
{\bf q}={\bf q'}\sqrt{2mnU(0)}/\hbar
\label{subst}
\end{equation}
in the integral [$U(0)$ is assumed to depend on $n$ in such a manner
that $nU(0)\to0$ when $n\to0$] and, then, taking the zero-density
limit in the integrand, for $n \to 0$ we find
\begin{equation}
I= 2\Lambda V(k),
\quad \Lambda=\ln(2c^{2}) mU(0)/(2\pi\hbar^2).
\label{lambda}
\end{equation}
Then, with the help of the Fourier transformation, Eq.~(\ref{Uk1}) reads
\[
\varphi(r)=1-\Lambda +\frac{m}{4\hbar^{2}}\int\d^{2}r'\,V(r')\varphi(r')
Y_{0}(q_{0}|{\bf r}-{\bf r}'|).
\]
Here the Fourier representation is used $Y_{0}(pr)= 4/(2\pi)^2
\mbox{P.P.} \int\d^2q\, \exp[i{\bf q}\cdot{\bf r}]/ (p^{2}-q^{2})$
for the cylindrical Bessel function of the second kind.
Since only the asymptotic of $\varphi(r)$ at $n\to0$ is of interest,
the linear integral equation for $\varphi(r)$ can be written as
\[
\varphi(r)=1-\Lambda +\frac{m}{2\pi\hbar^{2}}\!\int\d^{2}r'V(r')
\varphi(r')\ln(q_{0}|{\bf r}-{\bf r}'|e^{\gamma}/2),
\]
where the asymptotic $Y_{0}(z)=2\ln(ze^{\gamma}/2)/\pi+O(z^{2}\ln z)$
for $z\to0$ is used. Here $\gamma\simeq 0.5772$ stands for the Euler
constant, and $O(x)$ denotes terms of the order of $x$ or even higher.
It is seen from the resulted equation that, first, $\varphi(r)$ obeys
the Schr\"odinger equation (\ref{twobody}), and, second, its asymptotic
for $r\to\infty$ is
\begin{equation}
\varphi(r)\to 1-\Lambda+\ln(rq_{0}e^{\gamma}/2)mU(0)/(2\pi\hbar^{2}),
\label{bound1}
\end{equation}
which differs from that of Eq.~(\ref{bound}) only by the multiplication
factor $mU(0)/(2\pi\hbar^{2})$. Comparing Eq.~(\ref{bound1}) with
Eq.~(\ref{bound}) yields due to linearity of Eq.~(\ref{twobody})
\begin{eqnarray}
\varphi(r)&=&2u\varphi^{(0)}(r),
\label{phiru}\\
-\ln a&=&2\pi\hbar^{2}(1-\Lambda)/[mU(0)]+\ln(q_{0}e^{\gamma}/2),
\label{equ}
\end{eqnarray}
where we introduce the parameter $u$ by the definition
\begin{equation}
U(0)=\int\d^2r\,\varphi(r)V(r)=(4\pi\hbar^{2}/m)u.
\label{defu}
\end{equation}
With the help of Eq.~(\ref{lambda}) and the definition of $q_{0}$ (see
above), Eq.~(\ref{equ}) can be rewritten as
\begin{equation}
u=\delta (1+u\ln u), \quad \delta=-1/[\ln(na^{2}\pi)+2\gamma].
\label{udelta}
\end{equation}
Note that $\delta\to0$ at $n\to0$. As expected, the arbitrary
constant $c$ is cancelled and not involved in final
Eq.~(\ref{udelta}) and, hence, in the formula (\ref{phiru}) for
$\varphi(r)$. Equation (\ref{udelta}) has no solution for $u$ when
$\delta>1$ and has two positive ones when $\delta<1$ (i.e., when
$na^{2}<0.0369\ldots$). The solution with a greater value of $u$ should be
ignored because of its unphysical behaviour [$u\sim 1/(na^{2})$ at
$n\to0$]. An expansion for $u$ is obtained from Eq.~(\ref{udelta}) by
iterations
\begin{equation}
u=\delta+\delta^{2}\ln\delta +\delta^{3}\ln^{2}\delta +\delta^{3}
\ln\delta+O(\delta^{4}\ln^{2}\delta).
\label{uexp}
\end{equation}

Using Eq.~(\ref{nk}), one calculates the condensate depletion by means
of the substitution (\ref{subst}) upon integrating
\begin{equation}
\frac{n-n_{0}}{n}=\int\frac{\d^2q}{(2\pi)^{2}}n_{q}=u+\cdots.
\label{depletion}
\end{equation}
Thus, with the help of Eqs.~(\ref{phiru}), (\ref{defu}) and
(\ref{depletion}), one can rewrite Eqs.~(\ref{muasymp}) and
(\ref{grasymp}) as
\begin{eqnarray}
\mu&=&(4\pi\hbar^{2}n/m)u[1+u+\cdots],
\label{muexp} \\
g(r)&=&[\varphi^{(0)}(r)]^{2}4u^{2}[1+2u+\cdots].
\label{grexp}
\end{eqnarray}
Note that Eqs.~(\ref{phiru}) and (\ref{grexp}) are the short-range
approximation valid at $r\lesssim 1/\sqrt{n}$. For this reason, the
boundary condition (\ref{phi}) is not fulfilled for $\varphi(r)$ in
Eq.~(\ref{phiru}). In order to obtain the energy per particle
$\varepsilon$, we represent it in the form
$\varepsilon=(2\pi\hbar^{2}n/m)f(u)$ with an unknown function $f(u)$.
From Eq.~(\ref{udelta}) it follows that $n\partial u/\partial
n=u^{2}/(1-u)$, which, together with the thermodynamic relation
$\mu=\partial (\varepsilon n)/\partial n$, yields the differential
equation $u^{2}\d f/\d u + 2(1-u)f=2u(1-u^{2})$.  From this equation
and the initial condition $f(u=0)=0$ we derive
\begin{equation}
\varepsilon=(2\pi\hbar^{2}n/m)[u+u^{2}/2+O(u^{3})].
\label{eexpu}
\end{equation}
With Eq.~(\ref{uexp}), $\varepsilon$ can be expanded in the parameter
$\delta$
\begin{eqnarray}
\varepsilon&&=(2\pi\hbar^{2}n/m)\nonumber \\
           &&\times[\delta+\delta^{2}\ln\delta+\delta^{2}/2
+\delta^{3}\ln^{2}\delta+2\delta^{3}\ln\delta+O(\delta^{3})].
\label{eexpdelta}
\end{eqnarray}
One can see that all the low density expansions are series in the
dimensionless in-medium scattering amplitude $u$, which depends
ultimately on the density via Eq.~(\ref{udelta}) [and, hence,
(\ref{uexp})]; therefore, $u$ can be considered as {\it a parameter
of low density expansions in two dimensions}. The interaction energy
per particle is exactly related to the pair distribution function
(\ref{grexp})
\begin{eqnarray}
\varepsilon_{\mbox{\footnotesize int}}=\frac{n}{2}\!\int\d^{2}rV(r)g(r)
=\frac{2\pi\hbar^{2}n\alpha}{m}[u^{2}+2u^{3}+\cdots],
\label{eint}
\end{eqnarray}
where we put by definition
\begin{equation}
\alpha=\frac{m}{\pi\hbar^{2}}
\int\d^{2}r\,[\varphi^{(0)}(r)]^{2}V(r)
=\frac{2}{a}\frac{\partial a}{\partial \lambda}.
\label{alpha}
\end{equation}
In the latter equation we employ the theorem (\ref{varth}) with the
coupling constant $\lambda$ [$V(r)\to\lambda V(r)$, and $\lambda=1$
in final formulas].  Since $\varepsilon_{\mbox{\footnotesize int}}$
can be directly evaluated via the relation (\ref{eint}), our approach
takes accurately into account the short-range particle
correlations~\cite{our}.  Note that Eq.~(\ref{eint}) can also be
obtained from the Hellmann-Feynman theorem
$\varepsilon_{\mbox{\footnotesize int}}=\lambda \partial
\varepsilon/\partial \lambda$ with Eqs.~(\ref{udelta}),
(\ref{eexpu}), and (\ref{alpha}). Moreover, Eq.~(\ref{grexp}) can be
derived in the same manner varying the energy:  $g(r)=(2/n)\delta
\varepsilon/\delta V(r)$.  For the kinetic energy per particle
$\varepsilon_{\mbox{\footnotesize kin}}=\langle\sum
p^{2}_{i}\rangle/(2mN)$ we have
\begin{equation}
\varepsilon_{\mbox{\footnotesize kin}}=\varepsilon-
               \varepsilon_{\mbox{\footnotesize int}}
=\frac{2\pi\hbar^{2}n}{m}\Big[u+\Big(\frac{1}{2}-\alpha\Big)u^{2}+
                         \cdots\Big].
\label{ekin}
\end{equation}
It is seen from Eq.~(\ref{eexpu}), (\ref{eint}) and (\ref{ekin}) that
in the leading order, proportional to $nu$, the total energy is
purely kinetic. Thus, whatever a particular shape of the potential
$V(r)$, at sufficiently small densities the energy becomes mostly
kinetic. By contrast, in three dimensions
$\varepsilon_{\mbox{\footnotesize kin}} \simeq  2\pi \hbar^{2}bn/m$
and $\varepsilon_{\mbox{\footnotesize int}} \simeq 2 \pi
\hbar^{2}(a-b)n/m$ are of the same order, where $a$ is the 3D
scattering length and $b=a-\lambda\partial a/\partial
\lambda$~\cite{our} (except for hard-spheres when $a=b$, see
Ref.~\cite{lieb1}).

Let us discuss the nature of the Schick approximation
\begin{equation}
\varepsilon\simeq -2\pi\hbar^{2}n/(m\ln na^{2}),
\label{schickapp}
\end{equation}
which is Eq.~(\ref{eexpdelta}) in the lowest order in $n/\ln na^{2}$.
As the energy (\ref{schickapp}) in this order is purely kinetic, it
cannot in principle be represented as a sum of the interaction
energies of two particles over all pairs of bosons by analogy with
the weak-coupling 3D Bose gas. However, as in three dimensions, we
can start from Eq.~(\ref{mu}) and put $\mu\simeq
n\int\d^{2}r\,V(r)\varphi(r)$ with the {\it in-medium} pair wave
function $\varphi(r)$. It is clear that $\varphi(r)$ for $r\lesssim
r_{0}$ (here $r_{0}\sim 1/\sqrt{n}$ is of the order of the
correlation length) should be proportional to the wave function
$\varphi^{(0)}(r)$ of the two-body problem (\ref{twobody}):
$\varphi(r)\simeq C\varphi^{(0)}(r)$. The boundary condition
(\ref{phi}) can be fulfilled only {\it due to in-medium effects} for
$r\gtrsim r_{0}$; therefore, we can approximately put
$\varphi(r_{0})\simeq 1$ and use for $r_{0}\gg a$ the asymptotic
(\ref{bound}).  This leads to $C \simeq -2/\ln na^{2}$ and, by
Eq.~(\ref{iden}), yields $\mu\simeq -4\pi\hbar^{2}n/(m\ln na^{2})$
and, hence, Eq.~(\ref{schickapp}). The crucial difference in 3D case
is the boundary condition for the two-body problem
$\varphi^{(0)}(r)\to 1-a/r$ instead of Eq.~(\ref{bound}). In this
case the condition $\varphi(r_{0})\simeq 1$ leads to
$\varphi(r)\simeq \varphi^{(0)}(r)$ in the leading order, which
results in $\mu \simeq n \int\d^{3}r\,V(r)\varphi^{(0)}(r) =
4\pi\hbar^{2}na/m$.

Now one easily writes 2D Gross-Pitaevskii functional for the energy
using Eq.~(\ref{eexpu}) in the leading order
\begin{equation}
E[\phi]\!=\!\int\!\d^{2}r\bigg(\!
\frac{\hbar^{2}|\nabla\phi|^{2}}{2m}+
V_{\mbox{\footnotesize  ext}}({\bf r})|\phi|^{2}+
\frac{2\pi\hbar^{2}}{m}u|\phi|^{4}\!\bigg),
\label{GPfun}
\end{equation}
and 2D Gross-Pitaevskii equation
\begin{eqnarray}
i\hbar&& \partial \phi/\partial t=\delta E/\delta \phi^{*} \nonumber \\
&&=[-(\hbar^{2}/2m)\nabla^{2}+V_{\mbox{\footnotesize  ext}}({\bf r})]
\phi+(4\pi\hbar^{2}u/m)|\phi|^{2}\phi.
\label{GPeq}
\end{eqnarray}
Here $\phi=\phi({\bf r},t)=\langle\hat{\psi}({\bf r},t)\rangle$ is
the order parameter with the normalization $N = \int\d^{2}r \,
|\phi|^{2}$, and $u$ is given by Eq.~(\ref{udelta}) with
$n=|\phi|^{2}$. Upon varying in Eq.~(\ref{GPeq}), we neglect the
variation of $u$, for $|\phi|^{4}\delta u/\delta\phi^{*}\sim u^{2}|\phi|^{2}
\phi$. Note that the sum of the first and third terms in
Eq.~(\ref{GPfun}) corresponds to the {\it kinetic energy} of bosons
according to Eq.~(\ref{ekin}). Equations (\ref{GPfun}) and (\ref{GPeq})
are more exact than those of Ref.~\cite{shev}, based on the Schick
formula (\ref{schickapp}).

In conclusion, the expansions have been derived for the condensate
depletion (\ref{depletion}), the chemical potential (\ref{muexp}),
the pair distribution function (\ref{grexp}) for $r\lesssim
1/\sqrt{n}$, the total (\ref{eexpu}), interaction (\ref{eint}) and
kinetic (\ref{ekin}) energies. The energy expansion (\ref{eexpu})
leads to the Gross-Pitaevskii functional (\ref{GPfun}) and equation
(\ref{GPeq}). This work was supported by the RFBR grant 00-02-17181.


\appendix

\section{}
\label{scat}

In this appendix, a useful variational theorem is proved for 2D
scattering length. In this paper we deal only with the short-range
potentials which go to zero for $r\to \infty$ as $V(r)\to 1/r^{m}$
($m>2$), or even faster. The $s$-wave function corresponding to
relative motion of two particles with $p=0$ obeys the two-body
Schr\"odinger equation in the centre-of-mass system
\begin{equation}
-(\hbar^2/m)\nabla^2\varphi^{(0)}(r)+V(r)\varphi^{(0)}(r)=0
\label{twobody}
\end{equation}
with the following boundary conditions: first, $|\varphi^{(0)}(r)|<
\infty$ at $r=0$, and, second, for $r\to\infty$
\begin{equation}
\varphi^{(0)}(r)\to\ln(r/a).
\label{bound}
\end{equation}
Since the asymptotic is chosen to be real, the solution of
Eq.~(\ref{twobody}) is also real. The introduced positive quantity
$a$ is called 2D scattering length. Integrating
Eq.~(\ref{twobody}) and keeping in mind Eq.~(\ref{bound}) yield
\begin{equation}
2\pi\hbar^2/m=\int\d^{2}r\,V(r)\varphi^{(0)}(r).
\label{iden}
\end{equation}
Let us suppose that $V(r)$ is infinitesimally changed. Then, varying
Eq.~(\ref{twobody}), multiplying the obtained equation by
$\varphi^{(0)}(r)$, and carrying out the integration, one arrives at
the theorem
\begin{equation}
\frac{2\pi\hbar^2}{m}\frac{\delta a}{a}=
\int\d^{2}r\,[\varphi^{(0)}(r)]^{2}\delta V(r).
\label{varth}
\end{equation}

\end{document}